\documentstyle[aps,epsfig]{revtex}

\def\brho{{\mbox{\boldmath $\rho $}}}
\def\bomega{{\mbox{\boldmath $\Omega $}}}

\title{Positional, Reorientational and Bond Orientational Order in  DNA Mesophases} 

\author{ V. Lorman $^1$, R. Podgornik $^{2,3,4}$  and B. \v Zek\v s 
$^{5,3}$ }

\begin{document}


\twocolumn[\hsize\textwidth\columnwidth\hsize\csname@twocolumnfalse\endcsname

\maketitle

\noindent
{$^1$ Laboratoire de Physique Mathematique et Theorique,
Universite Montpellier II, F-34095 Montpellier, France}\\
\noindent 
{$^2$ Department of Physics, Faculty of Mathematics
and Physics, University of Ljubljana, SI-1000 Ljubljana, Slovenia}\\
\noindent
{$^3$ Department of Theoretical Physics, J.Stefan Institute, SI-1000 Ljubljana, 
Slovenia}\\
\noindent
{$^4$ LPSB/NICHHD, Bld. 12A  Rm. 2041,
National Institutes of Health, Bethesda, MD 20892-5626}\\
{$^5$ Institute of Biophysics, Medical Faculty, University of Ljubljana, SI-1000 Ljubljana, Slovenia}\\
\noindent

\begin{abstract}
\small
\noindent We investigate the orientational order of transverse
polarization vectors of long, stiff polymer molecules and their coupling
to bond orientational and positional order in high density mesophases. 
Homogeneous ordering of transverse polarization vector promotes
distortions in the hexatic phase, whereas inhomogeneous ordering
precipitates crystalization of the 2D sections with different
orientations of the transverse polarization vector on each molecule in
the unit cell.  We propose possible scenarios for going from the
hexatic phase, through the distorted hexatic phase to the crystalline
phase with an orthorhombic unit cell observed experimentally for the
case of DNA.

\end{abstract}
]

The microscopic structure of high density DNA mesophases is not only
of utmost importance for polymer science in general \cite{curoprudi}
but has a fundamental bearing also on topics as far removed from one
another as the nature of the phase diagram of magnetic vortex arrays
in type-II superconductors \cite{nelson1} or the structure of
genosomes used for DNA transfection in gene therapy related studies
\cite{lasic}.  By virtue of conventional wisdom at high densities DNA
should form crystals of hexagonal symmetry in the plane perpendicular
to the long axes of the molecules, with long range positional order,
and nematic (or better cholesteric) arrays with short range positional
order at densities intermediate between the crystalline and isotropic
phases.  There are however a few refinements one should add to this picture.

First of all it was found experimentally that DNA in fact forms
several crystalline phases \cite{langridge,grimm}.  At very high
densities $Li^+$-DNA makes a crystalline phase with orthorhombic
symmetry that implies a distorted hexagonal unit cell perpendicular to
the long axes.  The probable reason for this is that there are angular
frustrations between neighboring DNA molecules that make hexagonal
local symmetry energetically costly.  A similar situation can also be
encountered in frustrated spin systems, such as in
antiferromagnets on a triangular lattice or alkyl-chain systems
\cite{[13]}.  In all these cases distorting the hexagonal equilateral
into isosceles triangles could lower the angular part in the
interaction energy.  In this way two pairs of molecules are closer to
each other, maintaining optimal angles, while the third pair is
further apart and can be in a non-optimal configuration.  

In addition just below the crystalline phase DNA forms a line hexatic
phase with short range positional order and long range bond
orientational order \cite{pnas}.  The variation of the positional
correlation length as a function of DNA density shows that positional
order within this phase is more liquid-like (shorter correlation
length) the more DNA density is increased \cite{strey}.  This trend is
surprising and counterintuitive: one would expect the DNA array to
exhibit increasingly longer ranged positional order approaching the
crystalline phase where it becomes (ideally) infinite.  Again the
progressive disordering of DNA could be due \cite{strey} to increasing
angular frustration of molecules as they try to satisfy both the
positional and the angular constraints imposed by the interaction
potential.

The question of exactly how angular frustrations could affect
crystalline and bond-orientational order in DNA arrays will be
addressed in this contribution.  Angular frustrations of course
correspond to angle dependent terms in the interaction free energy
between DNAs.  Recent theoretical investigations
\cite{allahyarov} have indeed made it clear that at small
enough interaxial separations the interaction free energy depends
crucially on the mutual orientation of the two interacting molecules. 
This orientation can be specified in any 2D plane perpendicular to the
long axes, by giving {\sl e.g.} the position of the major groove of
one molecule with respect to the line joining a pair of them and of
the major groove of the other one with respect to the first one
\cite{allahyarov}.  This amounts to effectively defining a 2D vector
associated with each of the interacting molecules (we call it
transverse polarization vector, ${\bf p}$), perpendicular to their
long axes, and the interaction between them will depend on their
separation as well as their orientation described by their
transverse polarization vectors.  Basing our hypothesis loosely on
\cite{allahyarov} we assume that at low DNA density the
interaction does not depend on the mutual orientation of molecules, at
intermediate densities the interaction is minimal for parallel ${\bf
p}$ orientation while it is minimal for some finite angle between
${\bf p}$s for larger densities.

We assume that the molecules are stiff enough so that ordering of all
2D sections perpendicular to their long axes are the same.  We
consider consecutively two possible situations.  First at intermediate
densities the preferred angle between transverse polarization vectors
in all 2D sections is assumed to be zero.  The transverse
polarization order is thus homogeneous and we describe it with a 2D
constant vector $\bf p$, in the standard complex form ${\bf p} = \vert
p \vert e^{i \phi}$, where $\phi$ is the local angle between the
transverse polarization vector and some preferred axis.  In the
isotropic 2D liquid, hexatic order is associated with two symmetric
combinations of components of a 6th rank tensor.  In the complex form
the hexatic order parameter can be written in terms of the angle
$\theta$ between molecular bonds in the form $\psi= \vert \psi\vert
e^{i 6\theta}$, where $0 \leq \theta \leq \frac{2 \pi}{6}$.  The free
energy of hexatic ordering now depends only on ${\cal F}_{hex} = {\cal
F}_{hex}(\vert \psi\vert^2)$, while the free energy of polar
correlations can be written analogously as ${\cal F}_{pol} = {\cal
F}_{pol}(\vert p \vert^2)$.  The coupling of these two types of order
can be introduced through an interaction term ${\cal F}_{int} = {\cal
F}_{int}(\vert \psi\vert \vert p \vert^6 \cos{6(\phi - \theta}))$.  In
this picture the free energy of the isotropic liquid can become
unstable with respect to hexatic as well as polar order parameters.

\begin{figure}[htb]
\epsfxsize= 7.5 cm 
\centerline{\epsfbox{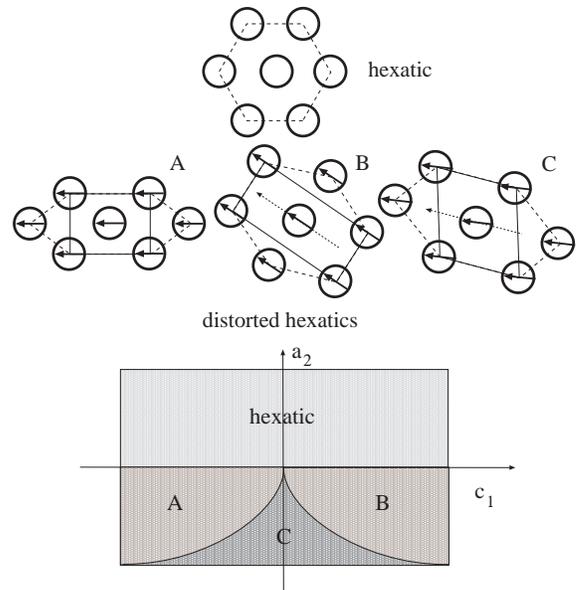}}
\caption{The 2D section of the phase diagram in coordinates $a_2$ and
$c_1$ for $a_1 < 0$, Eq.  \ref{eq:1}.  The bold lines mark second
order phase transition lines.  The order of the transverse
polarization vector can induce a transition into three types of
distorted hexatic phases, where the polar ordering is coupled to the
uniaxial distortions of the hexatic "unit cell" (there is no real unit
cell in a hexatic).  The hexatic order is indicated with dotted a
hexagon.  The directions of the transverse polarization vector in
phase (C) are in between (A) and (B) as indicated by the arrows on the
figure.  The resulting elementary unit cells are shown by bold lines}
\label{fig1}
\end{figure}

The total free energy ${\cal F} = {\cal F}_{hex} + {\cal F}_{pol} +
{\cal F}_{int} $ can now be written as
\begin{eqnarray}
{\cal F} &=& a_1 \vert \psi\vert^2 + b_1 \vert \psi\vert^4 + a_2 \vert p
\vert^2 + b_2 \vert p \vert^4 + \nonumber\\
&+& c_1  \vert \psi\vert \vert p \vert^6 \cos{6(\phi - \theta)} + \dots
\label{eq:1}
\end{eqnarray}
This free energy has five different stable solutions as obtained by
standard methods.  Since our main interest is in the transition from
the hexatic to the distorted hexatic phases, it is sufficient to
present only a typical cross-section of the phase diagram (described
in the coordinates $a_1,~a_2$ and $c_1$) in the region where the
isotropic liquid is already unstable, $a_1 < 0$.  In this cross-section with
coordinates $(a_2,~c_2)$ the phase diagram has a cusp-like form, Fig. 
\ref{fig1}.

The relevant stable solutions of Eq. \ref{eq:1} are in order of increasing density:\\
1. Isotropic 2D liquid: $\vert \psi\vert = 0$; $\vert p \vert =
0$\\
2. 2D hexatic phase: $\vert \psi\vert \neq 0$; $\vert p \vert =
0$\\
3. Three different distorted hexatic phases with the property: $\vert
\psi\vert \neq 0$; $\vert p \vert \neq 0$;
\begin{enumerate}
\item[(A)] One with: $\sin{6(\phi - \theta)} = 0$
    and $\cos{6(\phi - \theta)} = 1$, thus $\phi = \theta + \frac{
    \pi}{6} (2n + 1)$.  Obviously here the polar vectors are directed along
    the molecular bonds.  
\item[(B)] One with: $\sin{6(\phi - \theta)} = 0$
    and $\cos{6(\phi - \theta)} = -1$, thus $\phi = \theta + \frac{
    \pi n}{6}$.  Here the polar vectors make an angle $\frac{ \pi}{6}$
    with the bond directions.  
\item[(C)] Besides these two there also exists a
    lower-symmetry phase with a general angle between the polar vectors
    and the bond directions depending on external fields, such
    as density, and varying from $0^{\circ}$ to $30^{\circ}$.
\end{enumerate}
The point orientational symmetry in 2D planes of these phases is
$C_{6v}$ for the hexatic phase, $C_{s}$ for distorted hexatic phases
(A) and (B) and $C_1$ for the distorted hexatic phase (C).

In a typical case the transitions hexatic $\longrightarrow$ distorted
hexatic are of the second order.  The transition from the hexatic
phase to the low symmetry distorted hexatic phase (C) can be either
the succession of two indirect 2nd order transitions or one direct 1st
order transition, see Fig.  \ref{fig1}.  The distorted hexatic phases
A and B are isostructural and therefore at least in principle there
exists a continuous way to go between them without any phase
transition.  This scenario implies that the hexatic order parameter
$\vert \psi\vert$ decreases in these distorted hexatic phases until it
vanishes, goes through zero and reemerges with a different orientation
of the polar vector with respect to the local hexagonal axes.  More
formally this scenario implies that $\phi - \theta$ should go from
$0^{\circ}$ to $30^{\circ}$, a path which can be continuous only if it
crosses the point $\vert \psi\vert = 0$, where $\theta$ is not
defined.  If we assume that the hexatic order is strong enough and
vanishes at no point in the phase diagram, then the only way from
$\theta + \frac{\pi}{6}(2n + 1)$ to $\theta + \frac{ \pi n}{6}$
distorted hexatic is through two phase transitions.
The deformation of hexatic order (not shown in Fig.  \ref{fig1}) of
course follows the symmetry changes of the transverse polarization
vector.  It has to be uniaxial with one axis pointing in the direction
of the transverse polarization vectors of the molecules.  This can be
seen very easily if we introduce a parameter quantifying the local
distortion of the bond orientational (hexatic) field $\epsilon = \vert
\epsilon\vert~e^{i2\alpha}$, where $\epsilon$ is a combination of the
plane strain tensor $\epsilon_{ij}$ components: $ \epsilon =
\epsilon_{xx} - \epsilon_{yy} + 2i\epsilon_{xy};~\epsilon^* =
\epsilon_{xx} - \epsilon_{yy} - 2i\epsilon_{xy}$.

The coupling terms that do exist are between the transverse
polarization vector and the bond order distortion parameter and have
the form $ \vert p \vert^2 \vert \epsilon \vert \cos{2(\phi - \alpha
)}$.  This term induces the molecular bond distortion as a secondary
(improper) order parameter, when $p$ is different from zero.  In
principle we should also add the coupling term between the distortion
parameter and the hexatic order parameter that should be of the form $
\vert\epsilon\vert^3\vert \psi\vert \cos{3(\alpha - \theta)}$.  This
term shows that hexatic order in itself can not induce a distortion
and a coupling of the distortion parameter to the polar order is
{\it crucial}.

We now consider the case of larger densities where according to
theoretical estimates \cite{allahyarov} the preferred angle
between transverse polarization vectors of neighboring molecules is
non-zero.  The transverse polarization order is thus inhomogeneous and
we describe it with $\bf p(\brho)$, where $\brho = (x,y)$.  In this
case the distorted hexatic phases can give way to lower symmetry
phases characterized by a condensation of waves of the transverse
polarization vector.  The condensation of these modes invariably
induces also long range positional order pushing the system from a
distorted hexatic phase into a 2D crystalline phase with an
orthorhombic unit cell.

If the wave condensation starts from the distorted hexatic phase A (
or B ), then, in general, for transverse polarization modes, there
exist two qualitatively different choices of the wave vector ${\bf k}$
directions with respect to the symmetry plane of the distorted hexatic
order Fig.  \ref{fig1}.  The direction of the mode can either find
itself in the symmetry plane, in which case it can be described with a
single wave vector, or it can point in a general direction, when it
has to be described with the complete set of (two) directional wave
vectors.

The first case can evidently describe only one dimensional modes. 
Indeed, here the direction of the mode along y-axis, characterised by
$k_y$, is completely decoupled from the component along the x-axis. 
We thus obtain order parameter profiles which describe a
one-dimensional modulation of the transverse polarization
\textit{orientation}.  Note that the polarization {\bf p} is
already ordered in the homogeneously distorted hexatic phases ; then
the wave-like reorientation of {\bf p} in the (x,y)-plain is
described by the wave of the pseudo-vector perpendicular to this plain
and parallel to long molecular axes.  The angle \( \bomega (\brho ) \)
of the reorientation of {\bf p} with respect to its
direction in the homogeneously distorted hexatic phase (i.e. A-phase)
is given by the one-dimensional wave \( {\bomega }(\brho
)=\left| \bomega \right| \cos (k_{y}y+\beta ) \), where
\( \left| \bomega \right| \) is the module and \( \beta \) is the phase
of the corresponding order parameter (OP).

One of the possible realizations of the resulting ordered structure is
presented as 1 in Fig.  2.  Of course, the structure should depend on
the relation between the wavelength of the modulation and the average
distance between the molecules.  This relation cannot be obtained in
the framework of the phenomenological theory adopted in the present
work and depends on the constants of the microscopic interaction
between DNA molecules.  For definiteness, we present here (structure 1
in Fig.  2) the limit case, where the wave vector of the modulation is
taken to be \( k_{y}=\frac{2\pi }{b} \)\textbf{e}\( _{y} \).  Here \(
b \) is the long side of the orthorhombic ``local unit cell'' of the
homogeneously distorted hexatic phase A (Fig.  1) ; \textbf{e\( _{y}
\)} is a unit vector along the y-axis.  Then the angles of the
polarization {\bf p} directions in adjacent layers, taken
with respect to the y-axis are \( \left| \bomega \right| \)and \(
-\left| \bomega \right| \) respectively.

The only scalar invariant possible in the free energy descibing this
transition is $\vert \bomega \vert^2$ with the free energy itself assuming
the form
\begin{equation} 
{\cal F} = d_1 \vert \bomega \vert^2 + d_2 \vert \bomega \vert^4 + \dots.
\end{equation}
Typically this free energy describes one 2nd order phase transition at
$d_1 = 0$. In the x direction, being the direction perpendicular to
the direction of the condensed transverse polarization mode, the
system remains liquid after the phase transition, with continuous
translational symmetry.

In the second case the wave vector $\bf k$ is out of the symmetry
plane of the 2D section of the system.  The OP describing the
transverse polarization reorientation mode is in general four
dimensional.  Mathematically the irreducible representation has two
components but physically it has to have four components (if the waves
are to be real).  Therefore \( {\bomega _{1}}(\brho )=\left| \bomega
_{1}\right| e^{i\beta _{1}}e^{i{\bf k}_{1}\brho }\) ;
\( {\bomega _{2}}(\brho )=\left| \bomega _{2}\right| e^{i\beta
_{2}}e^{i{\bf k}_{2}\brho }\) ; \( {\bomega _{3}}(\brho
)={\bomega _{1}^{*}}(\brho ) \) ; and \( {\bomega _{4}}(\brho
)={\bomega _{2}^{*}}(\brho ) \) , where \textbf{k}\( _{1} \) and
\textbf{k}\( _{2} \) have the components \textbf{k}\( _{1} \)\(
=(k_{x,}\, k_{y}) \) and \textbf{k}\( _{2} \)\( =(-k_{x,}\, k_{y}) \)
along the symmetry axes of the ``unit cell'' of the homogeneously
distorted hexatic phase ; angles \( \left| \bomega _{1}\right| \) and
\( \left| \bomega _{2}\right| \) are two, in general independent wave
amplitudes.  The most general form of the resulting wave, which again
has to be real, is \( {\bomega }(\brho )=\sum
_{i}{\bomega _{i}}=(\left| \bomega _{1}\right| \cos ({\bf k}_{1}\brho
+\beta _{1})+\left| \bomega _{2}\right| \cos ({\bf k}_{2}\brho +\beta
_{2}))\) , with \( \beta _{1} \) and \( \beta _{2} \)
being the initial phases of the OP components that can be made zero by
an appropriate choice of the origin.

The free energy of the transverse polarization reorientation described
by this OP depends only on two independent invariants : \(
I_{1}=\left| \bomega _{1}\right| ^{2}+\left| \bomega _{2}\right| ^{2} \)
and \( I_{2}=\left| \bomega _{1}\right| ^{2}\left| \bomega _{2}\right|
^{2} \).  The corresponding free energy thus assumes the form
\begin{equation}
{\cal F} = d_1 I_{1} + d_2 I_{1}^2 + f_1 I_{2} + \dots
\label{freet}
\end{equation}
Apart from the distorted hexatic phases A, B or C with homogeneous
transverse polarization order $\vert \bomega_1\vert = 0$;~$\vert
\bomega_2\vert = 0$ already dealt with above, the minimization of this
free energy gives three different stable states:
\begin{enumerate}
\item[2a] $\vert \bomega_1\vert \neq 0$;~$\vert \bomega_2\vert = 0$ or
another domain where the roles of the fields $\vert \bomega_1\vert$
and $\vert \bomega_2\vert$ are reversed, \item[2b] $\vert
\bomega_1\vert = \vert \bomega_2\vert$, with either $\bomega_1 = \bomega_2$
or $\bomega_1 = -\bomega_2$, 
\item[2c] or the most general form $0 \neq \vert
\bomega_1\vert \neq \vert \bomega_2\vert \neq 0$.
\end{enumerate}
In what follows we shall limit ourselves to the distorted hexatic A and 
the phases with nonhomogeneous transverse polarization order that are 
derived from it, since the positional order in the crystalline phase of 
DNA strongly favors this type of distortion \cite{strey}.

\begin{figure}[htb]
\epsfxsize= 7.5 cm 
\centerline{\epsfbox{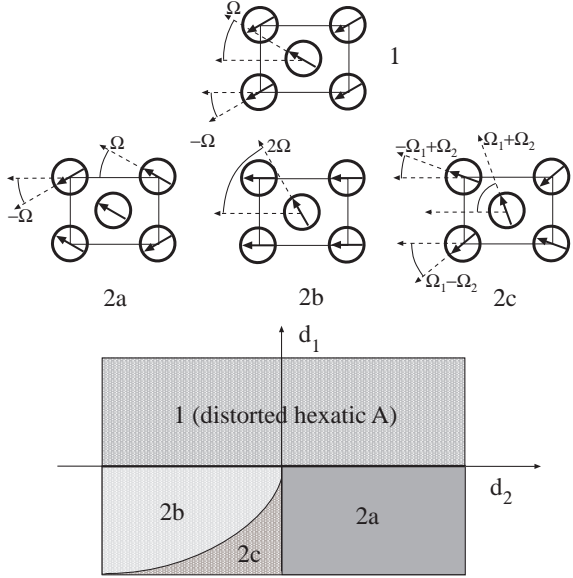}}
\caption{The 2D section of the phase diagram in coordinates $d_1$ and
$d_2$ for $f_1 > 0$, Eq.  \ref{freet}.  Distorted hexatic A, see Fig. 
\ref{fig1}, can give way to crystalline low symmetry phases with
condensed transverse polarization modes.  In (1) the transverse
polarization \textit{reorientation} mode condenses only in one
direction coinciding with one of the distorted hexatic ``unit cell''
vectors.  In phases (2) the angle \( \bomega \) of the polarization
reorientation is a sum of two modes with wave vectors out of the
symmetry plane of the 2D section, where the wave amplitudes \( \left|
\bomega _{i}\right| \) are in general independent.}
\label{fig2}
\end{figure}

The structure 2a, see Fig.  \ref{fig2}, has discrete translational
order only in one dimension with (as opposed to structure 1) the wave
vector making some general angle with the symmetry plane of the
uniformly distorted hexatic phase.  The transverse polarization
reorientation mode can be described with a simple form \(
{\bomega }(\brho )=\left| \bomega _{1}\right| \cos ({\bf k}_{1}\brho
) \).  The direction perpendicular to \textbf{k}\( _{1}
\) remains liquid.  Again for definiteness, we present the limit case
where \textbf{k}\( _{1} \) is taken to be \textbf{k}\( _{1} \)\(
=\frac{\pi }{a} \)\textbf{e}\( _{x} \)\( +\frac{\pi }{b}
\)\textbf{e}\( _{y} \) , \( a \) and \( b \) being the parameters of
the ``local unit cell'' of the phase A.
The structure 2b, see Fig.  \ref{fig2}, is represented by the sum of
two transverse polarization modes in the directions of ${\bf k}_1 $
and ${\bf k}_2 $ with equal amplitudes.  The polarization
reorientation mode thus assumes the form \( {\bomega }(\brho
)=\left| \bomega _{1}\right| (\cos ({\bf k}_{1}\brho )+\cos ({\bf k}_{2}\brho
)) \) .  Respecting the symmetry of the wave vectors we
present the limit case where \textbf{k}\( _{1} \) and \textbf{k}\(
_{2} \) are chosen in the form : \textbf{k}\( _{1} \)\( =\frac{\pi
}{a} \)\textbf{e}\( _{x} \)\( +\frac{\pi }{b} \)\textbf{e}\( _{y} \)
and \textbf{k}\( _{2} \)\( =-\frac{\pi }{a} \)\textbf{e}\( _{x} \)\(
+\frac{\pi }{b} \)\textbf{e}\( _{y} \) .  Note that the structures 1,
2a and 2b can be described using only one angular amplitude (see Fig. 
2).

The structure 2c, see Fig.  \ref{fig2}, presents the most general
structure of the ordered phase with two different directions of the
transverse polarization modes, ${\bf k}_1 $ and ${\bf k}_2 $ but also
with different amplitudes of the waves $\vert \bomega_1\vert \neq
\vert \bomega_2\vert$.  To present the limit structure we choose the
wave vectors \textbf{k}\( _{1} \) and \textbf{k}\( _{2} \) in the same
form as for the previous phase.  The symmetry is broken with respect
to the phases 2a and 2b : the directions of the transverse
polarizations of the molecules now make three different angles with
respect to the y-axis \( \left| \bomega _{1}\right| +\left| \bomega
_{2}\right| \), \( \left| \bomega _{1}\right| -\left| \bomega
_{2}\right| \) and \( -\left| \bomega _{1}\right| +\left| \bomega
_{2}\right| \).

The phase diagram corresponding to the free energy Eq.  \ref{freet}
has the form as presented on Fig.  \ref{fig2}.  All the transitions
between the distorted hexatic phase and modulated phases, as well as
between the modulated phases are of second order.

The existence of angular interactions among stiff, nematically
ordered, polymer molecules such as DNA, that depend on the orientation
of their transverse polarization vectors can introduce important
modifications into the phase diagram at intermediate and high
densities.  If the equilibrium transverse polarization order is
homogeneous (intermediate densities), corresponding to aligned
transverse polarization vectors on neighboring molecules, than the
hexatic phase becomes distorted with one axis of the local "unit cell"
pointing in the direction of the transverse polarization order.  This
distortion, that should grow as one approaches the hexatic - crystal
transition, could show up through a broader positional peak in X-ray
scattering, corresponding to domains of different distorted hexatic
directions, giving a very satisfactory explanation for observations in
DNA \cite{strey}.

If on the other hand, the equilibrium transverse
polarization order is inhomogeneous (higher densities) , corresponding
to non-zero angles between neighboring transverse polarization
vectors, the corresponding lattice becomes crystalline with long range
positional order either in one or both directions and with a deformed
hexagonal unit cell.  Thus instead of having six equivalent nearest
neighbors, the distorted hexagonal unit cell can give way to four
nearest neighbors at a non-zero angle (energetically more favourable)
between ${\bomega}$s, whereas the more distant molecules can have the
same direction of ${\bomega}$ (energetically less favourable).  This
case too could be associated directly with observations in
crystallised DNA arrays \cite{langridge}.  We are thus able to provide
a consistent interpretation for DNA phase behavior in a range of
densities between the crystalline and the cholesteric phases.

\end{document}